\documentstyle[12pt]{article}                                 

\begin{document}

\newcommand {\bea}{\begin{eqnarray}}
\newcommand {\eea}{\end{eqnarray}}
\newcommand {\be}{\begin{equation}}
\newcommand {\ee}{\end{equation}}

\title{
\begin{flushright}
\begin{small}
hep-th/9706071 \\
UPR/755-T \\
June 1997 \\
\end{small}
\end{flushright}
\vspace{1.cm}
Greybody factors for rotating black holes in four dimensions}

\author{Mirjam Cveti\v{c} and Finn Larsen\\
\small Department of Physics and Astronomy\\
\small University of Pennsylvania\\
\small Philadelphia, PA 19104 \\
\small e-mail: cvetic,larsen@cvetic.hep.upenn.edu
}
\date{ }
\maketitle

\begin{abstract}
We present the wave equation for a minimally coupled scalar field 
in the background of a rotating four-dimensional black hole that 
is parametrized by its mass, angular momentum, and four independent U(1) 
charges. The near horizon structure is identical to the 
five-dimensional case, and suggestive of an underlying description 
in string theory that is valid in the general non-extremal case. 
We calculate the greybody factors for the Hawking radiation. For
sufficiently large partial wave number the emission spectrum 
can be calculated for general non-extremal black holes and any particle 
energy. We interpret this spectrum in terms of a multi-body process 
in an effective string theory.
\end{abstract}  
                              
\newpage

\section{Introduction}
\label{sec:intro}
It is now widely accepted that a large class of extremal and 
near-extremal black holes can be described in string theory. 
The evidence for this correspondence falls in two broad categories: 
first, the counting of string states agrees with the degeneracies 
inferred from the Bekenstein-Hawking area 
law~\cite{sv1,callan96a} (also~\cite{sen95,structure,ct2}); 
and second, the rate and spectrum of Hawking radiation agrees in the two 
descriptions~\cite{callan96a,mathur,greybody}. The string theory 
side of the counting involves two distinct contributions to the 
entropy, from the right and left moving excitations of the string;
and similarly the emission spectrum of the string is characterized by 
two independent temperatures, because the initial state has quanta
in both sectors. The agreement requires that the classical geometry 
reproduces these duplications of thermodynamic variables; and this is 
indeed the case, at least in the so-called dilute gas regime.

The appearance of two independent sets of thermodynamic variables is 
related to the presence of two event horizons. For example the two 
contributions to the entropy are proportional to the sum and 
difference of the inner and outer horizon area, respectively. This 
geometric interpretion of the thermodynamic variables gives access to 
properties 
of the underlying microscopic theory that seems to be valid for 
{\it all} black holes. In a previous paper we explained  these 
results for the most general black holes in five dimensions~\cite{cl97a} 
(see also~\cite{cveticreview,fl97}). The purpose of the present paper 
is to present the corresponding calculations for a large class of 
rotating black holes in four dimensions. Our focus will be on the differences 
between the two cases; so a number of interpretive remarks will not 
be repeated.

The description of black holes in string theory is 
more involved in four dimensions than in five dimensions. 
In particular, the effective string models for the entropy are less 
rigorous~\cite{structure,strom96d,crr,kt,vf}; and the microscopic
foundation of the proposed exact BPS-degeneracy~\cite{dvv3} 
is less clear than the analogous one in five dimensions~\cite{sv1}. 
The regime of agreement between the emission spectrum of 
the string model and the Hawking radiation of semiclassical description 
is also more restrictive in four 
dimensions~\cite{klebanov96b,mathur97,strominger97a}. 

The interpretation of greybody factors is simplest in the regime where
the radiation is the result of a two-body process in the underlying 
string theory. In~\cite{cl97a} we found that for five dimensional black 
holes the rotational parameters can be tuned so that greybody factors 
retain their characteristic two-body form, without restrictions on 
the charges. Here we will find no comparable simplifications in four 
dimensions. Despite this limitation many cases remain where the 
radiation can be described as the result of a two-body process. An 
important example is the low energy limit of the radiation from the 
most general black holes.

New possibilities arise if we also consider multi-body processes 
in the effective string theory. It is particularly interesting to 
consider processes with higher partial wave number. 
In both four and five dimensions the partial wave number can be chosen
so that these processes can be described analytically in the 
classical theory for general non-extremal black holes, and even at high 
energy. The structure of the resulting emission rates has a characteristic 
multi-body form that can be interpreted in an effective string theory.

The paper is organized as follows. In sec.~\ref{sec:waveeq}, 
we present the wave equation for a minimally coupled scalar field in 
the background of a rotating black hole in four dimensions. 
We interpret the various terms. In sec.~\ref{sec:solution}, we 
solve the equation in various regions, and match these exact solutions 
to find approximate wave functions that are valid throughout. This leads 
to a general expression for the absorption cross-section. In 
sec.~\ref{sec:examples} the region of applicability of this result is 
determined, and we present examples. In sec.~\ref{sec:discussion} we 
discuss the microscopic interpretation of our classical calculations 
in terms of an effective string theory with emphasis on applications 
to general non-extremal black holes.

\section{The Wave Equation}
\label{sec:waveeq}
We consider a class of four-dimensional rotating black hole in 
toroidally compactified string theory. It is parametrized by the mass $M$, 
angular momentum $J$, and four independent $U(1)$ charges 
$Q_i$~\footnote{The most general black hole in four dimensions is
parametrized by five charges~\cite{ct2,hull96}. It has not yet been 
constructed.}. This black hole solution was given explicitly 
in~\cite{cy96b}; and
its non-rotating form appeared in~\cite{cystrings,hlm}. In special 
cases the black hole reduces to a solution of Einstein-Maxwell 
gravity. For example the Kerr-Newman black hole corresponds to all 
four charges equal. 

It is often convenient to parametrize the physical variables $M$,$Q_i$, 
and $J$ in terms of auxiliary quantities $\mu$,$\delta_i$, and $l$ 
defined as:
\bea
M &=& {1\over 2}\mu\sum_{i=1}^4 \cosh 2\delta_i~, \\
Q_i &=& {1\over 2}\mu \sinh 2\delta_i~~;~~~i=1,2,3,4~, \\
J &=& {1\over 2}\mu l~(\prod_{i=1}^4 \cosh\delta_i - 
\prod_{i=1}^4 \sinh\delta_i)~.
\label{eq:l12def}
\eea
The gravitational coupling constant in four dimensions 
is $G_4={1\over 8}$. In string units this corresponds to
$(2\pi)^6(\alpha^\prime)^4/V_6$.\footnote{The $m$ of~\cite{cy96b} 
is $m={1\over 4}\mu$ and also $l_{\rm there}={1\over 4}l_{\rm here}$.
The $r_0$ of~\cite{hlm} is $r_0={1\over 2}\mu$.}
With these conventions the areas $A_{\pm}$ of the inner and outer event 
horizons imply the entropies:
\bea
S_{\pm} &\equiv& {A_{\pm}\over 4G_N}=
2\pi[
{1\over 2}\mu^2 (\prod_i \cosh\delta_i+\prod_i \sinh\delta_i)
\pm
\sqrt{{1\over 4}\mu^4 (\prod_i \cosh\delta_i-\prod_i \sinh\delta_i)^2-J^2}]
\nonumber \\
&=&2\pi [
{1\over 2}\mu^2 (\prod_i \cosh\delta_i+\prod_i \sinh\delta_i)
\pm
{1\over 2}\mu\sqrt{\mu^2-l^2}(\prod_i \cosh\delta_i-\prod_i \sinh\delta_i)]~.
\label{eq:ent4d} 
\eea
In the string theory interpretation the two terms:
\bea
S_L &=& \pi\mu^2 (\prod_i \cosh\delta_i+\prod_i \sinh\delta_i), \\
S_R &=& {1\over 2}\mu\sqrt{\mu^2-l^2}
(\prod_i \cosh\delta_i-\prod_i \sinh\delta_i),
\eea
are the entropies of the left (L) and right (R) moving excitations.

We consider a minimally coupled scalar field propagating in the background 
of the black hole. The wave equation is the Klein-Gordon equation:
\be
{1\over\sqrt{-g}}\partial_\mu (\sqrt{-g}g^{\mu\nu}\partial_\nu \Phi) =0~.
\ee
It is convenient to introduce the dimensionless radial coordinate:
\be
x \equiv {r - {1\over 2}(r_{+}+r_{-})\over r_{+}-r_{-}}~,
\label{eq:xdef}
\ee
so that the outer and inner event horizons are located at
$x=\pm {1\over 2}$, respectively; and the asymptotic
space is at large $x$. The wave function is written in spherical 
coordinates as:
\be
\Phi\equiv\Phi_r(r)~\chi(\theta)
e^{-i\omega t+im\phi}~.
\ee
Then the radial part of the wave equation becomes:
\bea
&~&{\partial\over\partial x}(x^2-{1\over 4}){\partial\over\partial x}\Phi_r
+{1\over 4}[ x^2\Delta^2\omega^2 + 
xM\Delta\omega^2-4{\tilde\Lambda} \label{eq:geneq}
\\
&+&{1\over x-{1\over 2}}
({\omega\over\kappa_{+}}-m {\Omega\over\kappa_{+}})^2
-{1\over x+{1\over 2}}({\omega\over\kappa_{-}}-
m {\Omega\over\kappa_{+}})^2]\Phi_r = 0~. \nonumber
\eea
This equation is our main technical result\footnote{The main intermediate
step is to calculate the determinant of the metric. It turns
out that the result is simple: $-g=\Delta\sin^2\theta$, in the 
notation of~\cite{cy96b}.}. It is not more complicated than 
special cases that 
have been considered previously~\cite{klebanov96,mathur97,strominger97a},
but the present context is more general. In the following we 
explain the notation and interpret the various terms.

The variable $\Delta$ is defined as:
\be
\Delta = 2(r_{+} - r_{-}) = \sqrt{\mu^2 - l^2} = \beta_H^{-1}S~.
\label{eq:deltadef}
\ee
($T_H=\beta^{-1}_H$ is the Hawking temperature.) 
The first equation, and the definition of the radial variable $x$,
ensures that the wave equation reduces at large $x$ to the Klein-Gordon 
equation in {\it flat} space, as it should. Accordingly the term
${1\over 4}x^2 \Delta^2\omega^2$, dominant at large $x$, can be 
interpreted as the energy of the scalar field at infinity. 

At large $x$ the mass-term ${1\over 4}xM\Delta\omega^2$ is suppressed 
relative to the energy of the scalar field at infinity
by one power of $x\propto r$, as expected for a long 
range Coulomb type interaction in four dimensions. 

The effective angular momentum barrier $\tilde{\Lambda}$ is the term 
that is suppressed by $x^2\propto r^2$ for large $x$. It is:
\be
{\tilde\Lambda} = \Lambda - {1\over 16}\mu^2\omega^2 (1 + \sum_{i<j}
\cosh 2\delta_i \cosh 2\delta_j)~,
\ee
where $\Lambda$ is the eigenvalue of the operator:
\be
{\hat \Lambda} = -{1\over\sin\theta}{\partial\over\partial\theta}\sin\theta
{\partial\over\partial\theta} - {1\over\sin^2\theta}
{\partial\over\partial\phi^2}-{1\over 16}l^2 \omega^2 \cos^2\theta~,
\label{eq:anglap}
\ee
which is simply the angular Laplacian in four flat dimensions, except
for the last term which reflects the rotation of the black hole background.
 
The terms that are most important for the microscopic interpretation
are the horizon terms at $x=\pm {1\over 2}$. 
Their form is such that $\kappa_{\pm}$ 
are the physical surface accelerations at the inner and outer 
horizons~\cite{cl97a}. They are given by:
\bea
{1\over\kappa_{\pm}} 
&=&
{{1\over 4}\mu^3 (\prod_i \cosh^2\delta_i-\prod_i \sinh^2\delta_i)\over
\sqrt{{1\over 4}\mu^4 (\prod_i \cosh\delta_i-\prod_i \sinh\delta_i)^2-J^2}}
\pm
{1\over 2}\mu (\prod_i \cosh\delta_i-\prod_i \sinh\delta_i)
\nonumber \\
&=&
{\mu^2\over 2\sqrt{\mu^2-l^2}} 
(\prod_i \cosh\delta_i+\prod_i \sinh\delta_i)
\pm
{1\over 2}\mu (\prod_i \cosh\delta_i-\prod_i \sinh\delta_i)~.
\eea
Similarly, $\Omega$ is the angular velocity at the outer event horizon,
given by\footnote{The angular velocity at the inner event horizon
is given through
${1\over\kappa_{-}}\Omega_{-}={1\over\kappa_{+}}\Omega$. The appearance
of $\kappa_{+}$ in the inner horizon term is due to the convention that
$\Omega$ is measured at the outer horizon.}:
\be
{1\over\kappa_{+}}\Omega ={J\over
\sqrt{{1\over 4}\mu^4 (\prod_i \cosh\delta_i-\prod_i \sinh\delta_i)^2-J^2}}
= {l\over\sqrt{\mu^2-l^2}}~.
\ee
These expressions for $\kappa_{\pm}$ and $\Omega$ agree with those
that follow from the entropy (eq.~\ref{eq:ent4d}), by use of thermodynamic 
relations.

\section{Calculation of Absorption Cross-section}
\label{sec:solution}
The wave equation cannot in general be solved exactly. However we can 
find solutions valid in the asymptotic region and match them with 
solutions valid in the horizon region to find approximate wave functions 
that apply throughout spacetime. The calculations follow previous 
work closely~\cite{greybody,mathur97,strominger97a,dowker,hawking97}.

\paragraph{The angular equation:}
Consider first the angular Laplacian 
$\hat{\Lambda}$ (eq.~\ref{eq:anglap}). The eigenfunctions are the 
spheroidal functions (see {\it eg}.~\cite{abramowitz}):
\be
\chi(\theta)=S_{mn}({l\omega\over 4},\cos\theta)~.
\ee
The corresponding eigenvalues $\Lambda$ are labelled by the orbital 
angular momentum $n$ and the azimuthal quantum number $m$. They can be 
represented by a power series in $({l\omega\over 4})^2$ as:
\be
\Lambda (n,m) = n(n+1) + {1\over 2} 
[1 - {(2m-1)(2m+1)\over (2n-1)(2n+3)}]({l\omega\over 4})^2+\cdots 
\ee

\paragraph{The asymptotic region:}
Next consider the radial equation. At large $x\gg 1$ we omit the 
horizon terms so that the radial wave 
equation becomes:
\be
{\partial\over\partial x}x^2{\partial\over\partial x}\Phi_\infty
+{1\over 4}[x^2\Delta^2\omega^2+xM\Delta \omega^2-
4{\tilde\Lambda}]\Phi_\infty
= 0~.
\ee
Denoting the solutions in this region $\Phi_\infty^{\pm}$ we find:
\be
\Phi_\infty^{\pm} = x^{-{1\over 2}\pm \zeta} 
e^{-{i\over 2}\Delta\omega x} C_{\pm}
M_K({1\over 2}\pm\zeta-{i\over 4}M\omega,1\pm 2\zeta,i\Delta\omega x)~,
\label{eq:infwavefct}
\ee
where the function $M_K$ is Kummers function (see e.g.~\cite{abramowitz}), 
the parameter $\zeta$ is\footnote{The parameter $\xi$ of~\cite{cl97a} 
can be introduced as $\xi={1\over 2}+\zeta$. The present notation gives 
more symmetric formulae.}:
\be
\zeta = \sqrt{{1\over 4}+\tilde{\Lambda}}~,
\ee
and the normalization constants $C_{\pm}$ are:
\be
C_{\pm} = {1\over 2}(\Delta\omega)^{{1\over 2}\pm\zeta} 
e^{-{1\over 8}\pi M\omega} ~ 
{|\Gamma({1\over 2}\pm\zeta+{i\over 4}M\omega )|\over
\Gamma(1\pm 2\zeta)}~.
\ee
The wave functions eq.~\ref{eq:infwavefct} are generalizations of 
the Coulomb functions to non-integer angular momentum. They are normalized
so that:
\be
\Phi^{\pm}_\infty \sim x^{-{1\over 2}\pm\zeta}C_{\pm}~,
\label{eq:phiplus}
\ee
for $\Delta\omega x\ll 1$ and:
\be
\Phi^{\pm}_\infty \sim {1\over x}\sin ({1\over 2}\Delta\omega x - 
{1\over 4}M\omega\log x + {\rm const})~,
\label{eq:phiinf2}
\ee
for $\Delta\omega x\gg 1$.

\paragraph{The horizon region:}
In the horizon region we ignore the energy of the scalar field at infinity
and also the Coulomb type screening, due to the mass of the black hole. 
Denoting by $\Phi_0$ the radial wave function in this regime
the radial equation becomes:
\be
{\partial\over\partial x}(x^2-{1\over 4}){\partial\over\partial x}\Phi_0
-{\tilde\Lambda}+{1\over x-{1\over 2}}~{1\over 4}
({\omega\over\kappa_{+}}-m{\Omega\over\kappa_{+}})^2
-{1\over x+{1\over 2}}~{1\over 4}({\omega\over\kappa_{-}}-
m {\Omega\over\kappa_{+}})^2]\Phi_0 = 0
\ee
The form of this equation is identical to the corresponding one for 
the most general black holes in five dimensions~\cite{cl97a}. 
This is in harmony with the intuition that the near horizon terms 
express universal physics related to the underlying microscopic 
structure. In particular the $SL(2,R)_L\times SL(2,R)_R$ symmetry of the
horizon region, exhibited for the five-dimensional case in~\cite{cl97a},
carries over to four dimensions.

The solution relevant in the following has only an infalling 
component at the outer horizon. Taking the azimuthal quantum number
$m=0$ for typographical simplicity it is\footnote{We take $m=0$ in
the remainder of the paper. It can be restored by the replacements
$\beta_{R,L}\omega/2 \rightarrow \beta_{R,L}\omega/2 - m\beta_H\Omega$
and $\beta_H\omega\rightarrow \beta_H\omega-m\beta_H\Omega$.}:
\be
\Phi_0^{\rm in}=
({x-{1\over 2}\over x+{1\over 2}})^{-{i\beta_H\omega\over 4\pi}}
(x+{1\over 2})^{-{1\over 2}-\zeta} 
F({1\over 2}+\zeta-i{\beta_R\omega\over 4\pi},
{1\over 2}+\zeta-i{\beta_L\omega\over 4\pi},
1-i{\beta_H\omega\over 2\pi},{x-{1\over 2}\over x+{1\over 2}})~.
\label{eq:phiin0}
\ee
We introduced the inverse Hawking temperature 
$T^{-1}_H=\beta_H= {2\pi\over\kappa_{+}}$ and the corresponding
right (R) and left (L) components:
\bea
\beta_R &=& {2\pi\over\kappa_{+}}+{2\pi\over\kappa_{-}} 
={2\pi\mu^2\over\sqrt{\mu^2-l^2}}
(\prod_i \cosh\delta_i+\prod_i\sinh\delta_i)~,
\\
\beta_L &=& {2\pi\over\kappa_{+}}-{2\pi\over\kappa_{-}} =
2\pi\mu (\prod_i \cosh\delta_i-\prod_i \sinh\delta_i)~.
\eea
A linearly independent solution can be chosen as the purely outgoing 
wave, given by time reversal $\omega\rightarrow -\omega$. 
The asymptotic behavior of $\Phi^{\rm in}_0$ at large $x$ can be 
extracted, by using the modular properties of the hypergeometric 
function $F$. It is:
\be
\Phi_0^{\rm in}\sim x^{-{1\over 2}-\zeta}
{\Gamma(1-i{\beta_H\omega\over 2\pi})\Gamma(-2\zeta)\over
\Gamma({1\over 2}-\zeta-i{\beta_L\omega\over 4\pi})
\Gamma({1\over 2}-\zeta-i{\beta_R\omega\over 4\pi})}
+x^{-{1\over 2}+\zeta}
{\Gamma(1-i{\beta_H\omega\over 2\pi})\Gamma(2\zeta)\over
\Gamma({1\over 2}+\zeta-i{\beta_L\omega\over 4\pi})
\Gamma({1\over 2}+\zeta-i{\beta_R\omega\over 4\pi})}~.
\label{eq:phiininf}
\ee

\paragraph{}

It is now straightforward to calculate the 
absorption cross-section.
We take $A_\infty\Phi^{+}_\infty$ as the wave function in the asymptotic 
region, and in some intermediate ``matching region'' we identify this with 
$A_0\Phi_0^{\rm in}$, using eqs.~\ref{eq:phiplus} and~\ref{eq:phiininf}.
Postponing the justification of this procedure to the subsequent
section, the ratio of amplitudes becomes:
\be
|{A_0\over A_\infty}|= {1\over 2} (\Delta\omega)^{{1\over 2}+\zeta}~
e^{{\pi\over 8}M\omega}|\Gamma({1\over 2}+\zeta + {i\over 4}M\omega)|
~|{\Gamma({1\over 2}+\zeta-i{\beta_L\omega\over 4\pi})
\Gamma({1\over 2}+\zeta-i{\beta_R\omega\over 4\pi})
\over\Gamma(2\zeta)\Gamma(2\zeta+1)
\Gamma(1-i{\beta_H\omega\over 2\pi})}|~.
\ee
 From eq.~\ref{eq:phiinf2} we find the flux factor:
\be
{\rm flux}= {1\over 2i}(\bar{\Phi} r^2\partial_r\Phi-{\rm c.c.})
=  {\Delta^2\omega\over 16}|A_\infty|^2~,  
\ee
at large distances; and from eq.~\ref{eq:phiin0} we find 
the flux factor:
\be
{\rm flux}= {1\over 2i}(\bar{\Phi}r^2\sqrt{g^{rr}}\partial_r\Phi-{\rm c.c.})
={\beta_H\omega\Delta\over 8\pi}|A_0|^2 ~,  
\ee
at the horizon. The effective two dimensional transmission coefficient 
$|T_n|^2$ is the ratio of these two fluxes. Using a standard relation
from scattering theory we present the result as the absorption 
cross-section of the $n$th partial wave:
\bea
\sigma^{(n)}_{\rm abs}(\omega) &=& {\pi (2n+1)\over\omega^2}|T_n|^2 
\nonumber \\
&=& A~(2n+1)(\Delta\omega)^{2\zeta-1}
|{\Gamma({1\over 2}+\zeta-i{\beta_L\omega\over 4\pi})
\Gamma({1\over 2}+\zeta-i{\beta_R\omega\over 4\pi})
\over\Gamma(2\zeta)\Gamma(2\zeta+1)\Gamma(1-i{\beta_H\omega\over 2\pi})}|^2
\times \nonumber \\
&\times& 
e^{{\pi\over 4}M\omega}|\Gamma({1\over 2}+\zeta + {i\over 4}M\omega)|^2~.
\label{eq:partabs}
\eea
In the intermediate step we used 
${1\over 2}\beta_H\Delta = {1\over 2}S = A$ (which follows from
$G_N={1\over 8}$ and the last definition of $\Delta$ 
in eq.~\ref{eq:deltadef}.)

The absorption cross-section 
is almost identical to the corresponding expression in five 
dimensions~\cite{cl97a}; except for the presence of the last 
term. In the limit where the effective angular momentum barrier
$\tilde{\Lambda}$ vanishes we have $\zeta\sim{1\over 2}$; so this 
factor becomes:
\be
e^{{\pi\over 4}M\omega}|\Gamma(1+ {i\over 4}M\omega)|^2
={{1\over 2}\pi M\omega\over 1- e^{-{\pi\over 2} M\omega}}~.
\ee
The so-called Coulomb enhancement of the absorption is due to 
long range attractive interactions of Coulomb type. It is well-known 
from {\it e.g.} nuclear physics. Note that it is unique to four dimensions 
where the gravitational potential falls off as $r^{-1}$.

\section{Estimates and Examples}
\label{sec:examples}
\subsection{Matching on constant potential}
We now determine the regime of validity of the absorption cross-section 
eq.~\ref{eq:partabs}. The most generous condition arises when the
matching region can be chosen so that the effective angular momentum barrier 
$\tilde{\Lambda}$ dominates over all other terms. This is matching on 
a {\it constant} potential, in the parlance of~\cite{cl97a}. 
The logic is that the asymptotic wave function can be used at large $x$, 
the horizon wave function can be used at small $x$, and both are valid in 
the matching region. 

The precise condition is that there is a range of large $x$ so that:
\be
1\ll x~~;~
x^2 \Delta^2 \omega^2 \ll |\tilde{\Lambda}|~~;~
xM\Delta\omega^2 \ll |\tilde{\Lambda}|~~;~
{\beta_R \beta_L \omega^2\over x}\ll |\tilde{\Lambda}|
\ee
The necessary and sufficient conditions are that either:
\be
M\omega\gg |{\tilde\Lambda}|^{1\over 2}~~;~
\beta_R \beta_L M\omega^4\Delta\ll |{\tilde\Lambda}|^2  ~~;~
M\omega^2\Delta\ll |{\tilde\Lambda}| ~, 
\label{eq:const2}
\ee
or: 
\be
M\omega \tilde{<} ~|{\tilde\Lambda}|^{1\over 2} ~~;~
\beta_R \beta_L\omega^3 \Delta\ll 
|{\tilde\Lambda}|^{3\over 2}~~;~
\Delta\omega \ll |{\tilde\Lambda}|^{1\over 2} ~. 
\label{eq:const1}
\ee
In the S-wave ${\tilde\Lambda}\propto\omega^2$; so in this case
all the conditions are independent of frequency. This is reasonable 
because in this case the only frequency dependence of the potential 
is an overall factor that cannot influence the relative size of the terms.

In the calculation of the absorption cross-section we also assumed 
that the $x^{-{1\over 2}+\zeta}$ term dominates over the 
$x^{-{1\over 2}-\zeta}$ term in the matching region. The large value of 
the matching $x$ is sufficiently to ensure this, provided $\zeta>0$. 
In the regime where $\zeta$ is imaginary a closely related calculation
is valid. (It is given in the appendix of~\cite{mathur97}.)

\paragraph{Higher partial waves:}
The constant term in the potential is the effective angular momentum 
barrier. Thus an important example is that of higher partial waves. In 
particular, for any given background and perturbation frequency,
the partial wave number can be chosen so that ${\tilde\Lambda}\sim n^2$; 
and moreover so that all the conditions eq.~\ref{eq:const1} are satisfied. 
For natural frequencies $\omega\sim\beta_H^{-1}$ and generic black 
holes ($\delta_i\sim 1$ and $l\sim\mu$) it is sufficient that $n\gg 1$. 
This is interesting because the higher partial waves can be understood 
microscopically from multi-particle 
processes~\cite{strominger97a,mathur97b,gubser}. We will consider this 
example further in the discussion in sec.~\ref{sec:discussion}.

\paragraph{Two large boosts:}
This case is the S-wave, with two large boosts 
$\delta\sim\delta_i\gg 1$, and the other two boosts of order unity. 
A background angular momentum of order $l\sim\mu$ may be included.
The inverse temperatures are comparable:
$\beta_R\sim\beta_L\sim\mu e^{2\delta}$; and
the other parameters are $M\sim\mu e^{2\delta}$,
$|{\tilde\Lambda}|\sim\mu^2 \omega^2 e^{4\delta}$, and $\Delta\sim\mu$.
The conditions eq.~\ref{eq:const1} are satisfied for all frequencies.

\paragraph{}
We note that the case of S-wave and one large boost violates the 
conditions, even if the freedom to tune the angular momentum is employed.
In this case the absorption cross-section cannot be found in closed form.

\subsection{Matching on vanishing potential}

The calculation of Maldacena and Strominger~\cite{greybody} employed
matching on a {\it vanishing} potential. This is in general more 
restrictive than the matching on a constant potential, but it also 
provides a clearer distinction between the horizon terms, of presumed
microscopic importance, and the long range fields. In this case
there must be a range of large $x$ so that {\it all} the potential 
terms can be neglected:
\be
1\ll x~~;~
x^2 \Delta^2 \omega^2 \ll 1 ~~;~
xM\Delta\omega^2 \ll 1 ~~;~
|\tilde{\Lambda}| \ll 1 ~~;~
{\beta_R \beta_L \omega^2\over x} \ll 1~.
\ee
The necessary and sufficient conditions are that either:
\be
M\omega\gg 1~~;~
|\tilde{\Lambda}| \ll 1 ~~;~
\beta_R \beta_L M\Delta\omega^4 \ll 1 ~~;~
 M\Delta\omega^2 \ll 1~, 
\ee
or: 
\be
M\omega \tilde{<} 1 ~~;~
|\tilde{\Lambda}| \ll 1 ~~;~
\beta_R \beta_L \Delta\omega^3 \ll 1 ~~;~
 \Delta\omega \ll 1~. 
\label{eq:matchcond}
\ee
We have not found any significant example that takes advantage of the 
first conditions, and a similar comment applies to scattering that is not 
in the S-wave; so these possibilities will be disregarded.
Then the conditions eq.~\ref{eq:matchcond} are low energy conditions. 

Matching on a vanishing potential automatically implies
$\zeta\sim{1\over 2}$; so the general absorption cross-section 
eq.~\ref{eq:partabs} simplifies dramatically:
\bea
\sigma^{(0)}_{\rm abs}(\omega)
&=& A |{\Gamma(1-i{\beta_L\omega\over 4\pi})
\Gamma(1-i{\beta_R\omega\over 4\pi})
\over\Gamma(1-i{\beta_H\omega\over 2\pi})}|^2
~e^{{\pi\over 4}M\omega}|\Gamma(1+ {i\over 4}M\omega)|^2 \\
&=&
A~{\beta_L {\omega\over 2}\beta_R {\omega\over 2}\over\beta_H\omega}~
{e^{\beta_H \omega}-1\over 
(e^{\beta_L {\omega\over 2}}-1)(e^{\beta_R {\omega\over 2}}-1)}
~{{1\over 2}\pi M\omega\over 1- e^{-{\pi\over 2}M\omega}}
\eea
The Coulomb enhancement factor will play no role in our discussion.
Next we consider some examples:

\paragraph{Low energy:}
All conditions are automatically satisfied when the wave length 
$\lambda\sim\omega^{-1}$ is larger than all other scales in the problem. 
In this case:
\be
\sigma_{\rm abs}(\omega\rightarrow 0) = A~.
\ee
This is the universal low energy absorption cross-section 
(see~\cite{gibbons96} and references therein.) 

\paragraph{Dilute gas regime:}
In this case three boosts are large $\delta\sim\delta_i\gg 1$ but the 
last one is of order unity~\cite{klebanov96}. An angular momentum of order 
$l\sim\mu$ may be included. The inverse temperatures are comparable
$\beta_R\sim\beta_L\sim \mu e^{3\delta}$; and the other parameters are
$M\sim\mu e^{2\delta}$, 
$|{\tilde\Lambda}|\sim \mu^2 \omega^2 e^{4\delta}$, and $\Delta\sim\mu$.
The effective angular momentum barrier and the horizons terms
give the strongest condition on the frequency, namely 
$\mu\omega e^{2\delta}\ll 1$. However this is still consistent 
with the interesting regime
$\beta_R\omega\sim\beta_L\omega\sim 1$.

\paragraph{Near BPS black hole:}
Here all boosts are large $\delta\sim\delta_i\gg 1$. An angular
momentum of order $l\sim\mu$ may be included. There is a hierarchy 
between the temperatures: $\beta_L\sim\mu e^{2\delta}$ and 
$\beta_R\sim\mu e^{4\delta}$; and the other parameters are
$M\sim\mu e^{2\delta}$,
$|{\tilde\Lambda}|\sim\mu^2 \omega^2 e^{4\delta}$,
and $\Delta\sim\mu$. The horizon terms and the effective
angular momentum barrier both give the condition 
$\mu\omega e^{2\delta}\ll 1$ on the frequency. This is consistent with 
the structure in the range $\beta_R\omega\sim 1$ but not the other 
interesting range $\beta_H\omega\sim\beta_L\omega\sim 1$.

\paragraph{Near extremal Kerr-Newman:}
The angular momentum is tuned so that $\mu^2-l^2=\mu^2\epsilon^2\ll\mu^2$,
but the charges are left general. In this limit
there is a hierarchy between the temperatures: 
$\beta_L\sim\mu$ and $\beta_R\sim\mu\epsilon^{-1}$; and
$|{\tilde\Lambda}|\sim\mu^2\omega^2$, $M\sim\mu$, and
$\Delta\sim\mu\epsilon$. The strongest condition on the frequency,
from the horizon terms and the effective angular momentum barrier,
is $\mu\omega\ll 1$. This is analogous to the near BPS case: the 
region $\beta_R\omega\sim 1$ can be probed, but 
$\beta_H\omega\sim\beta_L\omega\sim 1$ cannot.

\paragraph{}

In five dimensions there are two angular momenta and it is possible to tune
them so that the inverse temperatures are large and 
comparable~\cite{cl97a}. In this
limit the calculation is sensitive to all boost parameters. However in
four dimensions the angular momentum appears only in the R sector. It 
plays no role in the estimates unless $l$ is tuned to be near $\mu$, and 
this implies $\beta_R\gg\beta_L$; so there is necessarily a hierarchy 
between the inverse temperatures. There is therefore no 
``rapidly spinning black hole'' regime in four dimensions. 

\section{Discussion}
\label{sec:discussion}
In the region where matching on a vanishing potential can be justified,
the Hawking emission rate is\footnote{We ignore the Coulomb-type
factor. Note that in regimes where it gives an enhancement of the absorption
cross-section it gives a {\it reduction} of 
the emission rate. The naive application of detailed balance
that leads to eq.~\ref{eq:emrate} is modified by the presence of long 
range interactions.}:
\bea
\Gamma_{\rm em}^{(0)}(\omega) &=&\sigma_{\rm abs}(\omega)~
{1\over e^{\beta_H\omega}-1}~{d^4 k\over (2\pi)^4} 
\label{eq:emrate} \\
&=&
2\pi G_N~{\cal L}\omega~
{1\over (e^{\beta_L {\omega\over 2}}-1)(e^{\beta_R {\omega\over 2}}-1)}
~{d^4 k\over (2\pi)^4}~,
\label{eq:gammaem}
\eea
where:
\be
{\cal L}=
2\pi\mu^3 (\prod_{i=1}^4 \cosh^2\delta_i -\prod_{i=1}^4 \sinh^2\delta_i)~.
\label{eq:leff}
\ee 
The emission rate eq.~\ref{eq:gammaem} is identical to the two-body 
annhilation rate for quanta propagating on an effective string of length 
${\cal L}$~\cite{mathur}. From our classical
calculation we expect that this microcopic model of the emission 
process can be applied in the full range of validity (eq.~\ref{eq:matchcond})
of the two-body form of the emission rate (eq.~\ref{eq:gammaem}).
In particular, this includes the emission of very low energy quanta 
from {\it all} black holes; so in this sense we can model all black holes 
as effective strings, and their Hawking radiation as two-body processes.

However, it is only for special black holes that the two-body description 
can be used at the energies of typical Hawking particles; and only for these 
black holes is the emission spectrum eq.~\ref{eq:gammaem} applicable in 
the range where it has characteristic features. In four dimensions there 
is less flexibility than in five dimensions; so the characteristic 
two-body features are only significant in regimes that involve 
restrictions on the charges. 

Eventually we would like to have a microscopic description of
all processes, at least in principle. However, at this point it seems 
more rewarding to consider specific processes that go beyond the 
S-wave of a minimally coupled scalar 
field~\cite{cgkt,mathur97,strominger97a,mathur97b,gubser}. 
Higher partial waves are particularly interesting because, 
for sufficiently high partial wave number (eq.~\ref{eq:const1}), 
their absorption cross-section can be calculated in the classical
theory for all black holes and arbitrary 
frequencies\footnote{We ignore backreaction on the geometry and 
ultimately this sets on both frequency and partial wave number. 
Concerns have been raised that for this reason the microscopic 
description of very high partial waves may not be 
accurate~\cite{gubser}.}; and the result has a form that is characteristic 
of string theory\footnote{These statements are 
also true in the five dimensional case.}. The corresponding 
emission rate can be written\footnote{As before we suppress the Coulomb-type 
factor due to long range fields.}:
\bea
\Gamma_{\rm em}^{(n)}(\omega)
&=& ({8G_N {\cal L}\omega\over 2\pi})^{2n+1} 
(2n+1){(2\pi)^2(n!)^2\over 8(2n)!^2(2n+1)!^2} 
~
{\prod_{j=1}^n [({\omega\over 2})^2 + ({2\pi j\over\beta_L})^2 ]
\over (e^{\beta_L {\omega\over 2}}-1)} \times \nonumber \\
&\times &
{\prod_{j=1}^n [({\omega\over 2})^2 + ({2\pi j\over\beta_R})^2 ]
\over (e^{\beta_R {\omega\over 2}}-1)}
~{d^3 k\over (2\pi)^3}~.
\label{eq:nem} 
\eea
This expression depends only on quantities that have a microscopic 
interpretation: $\beta_{R,L}$ are the inverse temperatures of 
the right and left moving string excitations, ${\cal L}$ is the 
length of the effective string, and Newton's coupling constant $G_N$ 
is the $U$-duality invariant form of the string coupling. The angular 
momentum of the black hole background enters only through $\beta_R$. 
This is expected from a microscopic point of view, because here the 
introduction of angular momentum is implemented as a projection acting 
on the Hilbert space of the right movers. (In five dimensions there are 
two angular momenta and they enter through $\beta_R$ and $\beta_L$, 
respectively.)

In an effective string theory description the emission of a partial
wave with an angular momentum $n$ is dominated by an operator 
that has dimension $n+1$, both in the right and left moving sectors. 
It can be realized as a composite operator of $n+1$ free boson fields
(each boson can be traded for $2$ fermions). It is simplest to calculate 
the thermal phase space factors of the initial string state in the 
bosonic realization:
\be
I^{(n+1)}(\beta) = \int^\infty_{-\infty} 
\delta ({\omega\over 2}-\sum_{i=0}^{n} p^i)~
\prod_{i=0}^{n}{p^i dp^i\over e^{\beta p^i}-1} = 
 {1\over (2n+1)!}~{\omega\over 2}~
{\prod_{j=1}^n [({\omega\over 2})^2 + ({2\pi j\over\beta})^2 ]
\over (e^{\beta{\omega\over 2}}-1)}~.
\ee
(Useful relations are given in~\cite{gubser}.) The final state is the 
$n$th partial wave of a minimally coupled scalar field. Microscopically 
this corresponds a vertex operator of a scalar field with $n$ 
spacetime derivatives; so this gives a further 
frequency dependence $\omega^n$ in the amplitude, and therefore 
$\omega^{2n}$ in the rate. Finally, the normalization of the outgoing 
state gives a factor $\omega^{-1}$. Collecting the factors, we arrive at 
a microscopic interpretation of the frequency dependence of the emission 
rate eq.~\ref{eq:nem}. 

The dependence on coupling constant can be understood as follows: 
in the bosonic representation there are $n+1$ operators on both the 
right and left sides, and there is one outgoing state. The sphere 
amplitude with these $2n+3$ vertices has a factor of 
$g^{-2}g^{2n+3}=g^{2n+1}$; so the rate has a factor of 
$g^{4n+2}\propto G_N^{2n+1}$. Moreover, 
there is a supressed factor of ${\cal L}$ in the measure of 
each of the phase 
space integrals above, and the outgoing particle accounts for a factor 
of ${\cal L}^{-1}$; so the dependence on the effective string length 
becomes ${\cal L}^{2n+1}$. Dimensional analysis serves as a 
check on these arguments. Combining these results we account for 
eq.~\ref{eq:nem}, up to the numerical prefactor.

It would clearly be desirable to understand these arguments in the 
framework of a detailed microscopic model. Specifically the numerical 
prefactor of the emission rate should be calculable. However, already 
in its present form the analysis indicates a striking connection 
between the classical result and an effective string theory.

It is apparent from the preceding discussion that the emission rate 
eq.~\ref{eq:nem} is closely related to similar ones that appear in 
the context of near-extremal black holes. An effort to 
understand these cases microscopically, including prefactor, is
far advanced and our discussion is adapted from this 
context~\cite{strominger97a,mathur97b,gubser}\footnote{An important
issue in these calculations is the dependence on the volume of the 
compact space. Our approach is manifestly duality invariant; so it 
supports the resolution proposed in~\cite{mathur97b}.}. However, 
our result generalizes these limiting cases: (i) it treats 
all four charges independently, and no hierarchy between them is assumed, 
(ii) it is {\it not} limited to near-extremal black holes, 
(iii) it includes a background angular momentum, and 
(iv) there is no low energy requirement. The expression eq.~\ref{eq:nem}
is nevertheless of the form that has been considered previously. 
It is reasonable to assume that an effective string model can be devised 
that accounts for the details of the limiting cases; and then our 
classical result seems to indicate that the model will automatically 
apply also in general.

\vspace{0.2in} {\bf Acknowledgments:} 
This work is supported in part by DOE grant DOE-EY-76-02-3071, NSF
Career advancement award PHY95-12732 (MC), and NATO collaborative
grant CGR 949870 (MC).

\end{document}